\documentclass[11pt]{article}

\usepackage{amsmath}
\usepackage{amsthm}
\usepackage{amscd}
\usepackage{amsxtra}
\usepackage{upref}
\usepackage{amsfonts}
\usepackage{amssymb}
\usepackage{graphicx}
\usepackage{stmaryrd}
\usepackage{url}
\usepackage{cite}
\usepackage{pifont}
\usepackage{color}
\usepackage{hyperref}
\usepackage{todonotes}
\def\beq{\begin{equation}}
\def\eeq{\end{equation}}
\def\be{\begin{equation}}
\def\ee{\end{equation}}

\newcommand{\half}{{{\textstyle\frac{1}{2}}}}

\newcommand{\beqa}{\begin{eqnarray} }
\newcommand{\eeqa}{\end{eqnarray} }
\newcommand{\ba}{\begin{array}}
\newcommand{\ea}{\end{array}}

\newcommand{\SLf}{\mathrm{SL}(5)}

\def\cL{\mathcal{L}}

\def\tg{\tilde{g}}

\def\tv{\tilde{v}}
\def\tx{\tilde{x}}

\def\tm{\tilde{m}}
\def\tM{\tilde{M}}
\def\tpartial{\tilde{\partial}}

\def\tR{\tilde{R}}
\def\tH{\tilde{H}}
\def\txi{\tilde{\xi}}
\def\tC{\tilde{C}}
\def\tphi{\tilde{\phi}}
\def\cM{\mathcal{M}}
\def\tlambda{\tilde{\lambda}}

\numberwithin{equation}{section}

\begin{document}
\begin{titlepage}
\title{M-theory and Type IIB from a Duality Manifest  Action}
\author{\sc Chris D. A. Blair$^*$, Emanuel Malek$^*$ and Jeong-Hyuck Park$^{*,\dagger}$}
\date{}
\maketitle
\begin{center}
$^*$DAMTP, Centre for Mathematical Sciences, University of Cambridge, \\
Wilberforce Road, Cambridge CB3 0WA, United Kingdom. \\
$^\dagger$Department of Physics, Sogang University, Mapo-gu, Seoul 121-742,
Korea.\footnote{On sabbatical leave.}\\
~\\
\texttt{C.D.A.Blair@damtp.cam.ac.uk\quad E.Malek@damtp.cam.ac.uk \quad
Park@sogang.ac.kr}
~~~\\~\\
\end{center}
\begin{abstract}
We revisit the $\SLf$ U-duality manifest action constructed by Berman and Perry
in an extended spacetime. Upon choosing a four-dimensional
solution to the section condition constraint, the theory reduces to a four-dimensional truncation of
eleven-dimensional supergravity. In this paper, we show that the theory
contains more than this M-theory reduction. The section
condition also admits an $\SLf$ inequivalent three-dimensional solution, upon
which the action directly reduces to a three-dimensional truncation of type
IIB supergravity. 
We
also discuss the reduction to IIB$^*$ supergravity.
\end{abstract}
\thispagestyle{empty}
\end{titlepage}
\newpage
\section{Introduction}

The presence of string theory dualities leads to the idea of M-theory as a
unification of all superstring theories \cite{Hull:1994ys,Witten:1995ex}. These dualities map
between different theories, and allow one to identify individual string theories as limits of
M-theory. 

For instance, type IIA string theory can be viewed as M-theory compactified on a circle. In order to further make an identification with type IIB we have to compactify a second direction and use T-duality.
In this paper, working within the
context of the low energy effective theories, we will show how a recently
developed formulation that makes dualities
manifest allows one to directly identify M-theory/IIA and IIB within a single
framework. 

In doing so we draw inspiration from several different approaches to
understanding string theories and
their low energy actions
through dualities. The common theme of these approaches - reminiscent of the
introduction of
M-theory as a higher-dimensional theory allowing one to understand features of the original
superstring - is to view dualities as manifest symmetries of some larger, often geometric, structure,
which reduces to or contains the initial theory of interest. 

F-theory \cite{Vafa:1996xn} provides a realisation of IIB as a reduction from a
higher-dimensional theory. The extra dimensions geometrise the
$\mathrm{SL}(2)$ S-duality of IIB supergravity so that one has a
twelve-dimensional theory. There were
hints that this geometrisation could be extended to the U-dualities of type IIB \cite{Kumar:1996zx}. 

More recently, the introduction of generalised geometry
by Hitchin and Gualtieri \cite{Hitchin:2004ut,Gualtieri:2003dx} has
provided a framework in which one makes dualities manifest by extending the
tangent bundle $TM$. For
T-dualities 
\cite{Grana:2008yw,Coimbra:2011nw,Coimbra:2012yy}
one studies the extended tangent bundle $TM \oplus T^*M$, allowing one to
combine
diffeomorphisms and B-field gauge transformations. The extra factor $T^*M$
is needed to take into account the effects of string windings. A natural object
appearing on the extended tangent bundle is the generalised metric, originally
introduced in \cite{Duff:1989tf}, which combines the metric and $B$-field into a
T-duality covariant object.

By further adding appropriate bundles for D-brane wrappings one arrives at
generalised geometries that describe U-duality in type
II theories \cite{Grana:2009im,Aldazabal:2010ef,Grana:2011nb,Coimbra:2011ky}. In
this approach the IIA and IIB theories have to be treated differently, i.e.
they require
different choices of generalised geometries. Their extended tangent
bundles
are $TM \oplus T^*M \oplus \Lambda^{\mathrm{even/odd}} T^*M \oplus \cdots$,
with the even/odd exterior products of the
cotangent bundle corresponding to IIA/IIB. One can also extend this treatment to
M-theory
\cite{Hull:2007zu,Pacheco:2008ps,Coimbra:2011ky,Coimbra:2012af}, where one now
has wrappings of the M2 and M5 branes. The corresponding extended tangent bundle is $TM \oplus
\Lambda^2 T^*M \oplus \Lambda^5 T^*M \oplus \cdots$. One can see that in
this approach type IIA arises from the usual reduction of M-theory, whereas
a different extended tangent bundle is used for IIB.

In order to make the dualities manifest in the action, one can introduce extra
coordinates, as pioneered by Duff, Tseytlin and Siegel in the early 90s
\cite{Duff:1989tf,Duff:1990hn,Tseytlin:1990nb,Tseytlin:1990va,Siegel:1993xq,
Siegel:1993th}. The recent revival of these ideas has led to a T-duality
manifest rewriting of ten-dimensional
supergravity known as double field theory \cite{Hull:2009mi, Hull:2009zb,
Hohm:2010jy, Hohm:2010pp,Hohm:2010xe,
Jeon:2010rw, Jeon:2011kp,Jeon:2011cn, Jeon:2011vx, Jeon:2011sq,Jeon:2012kd,Jeon:2012hp,
Hohm:2011ex, Hohm:2011zr,Hohm:2011dv, Hohm:2011nu,Aldazabal:2013sca}. An
analogous programme
exists for M-theory and eleven-dimensional supergravity
\cite{Berman:2010is,Berman:2011cg,Berman:2011jh,Thompson:2011uw,Berman:2013eva,Godazgar:2013rja,
Park:2013gaj,Aldazabal:2013mya,Cederwall:2013naa}, and makes the U-dualities
manifest symmetries of the theory. In the extended theories, the fields do not
however depend on all of the coordinates. All physical fields and gauge
parameters have to satisfy the ``section condition'' which reduces their
dependence and allows one to return to the familiar unextended  theory. In fact,
imposing the section condition can be shown to be  equivalent to
``gauging'' the extended coordinate space~\cite{Park:2013mpa,Lee:2013hma}.

One might argue that due to the section condition, these extended theories are
locally equivalent to the description of generalised geometry
\cite{Hull:2007zu,Grana:2008yw,Pacheco:2008ps}
where no extra coordinates are introduced. In this paper we will show that the
U-duality extended theory encompasses more than just the generalised geometry
for M-theory from which it was constructed. It contains a new solution to the
section condition which gives rise directly to the IIB theory. 
We will show how this works explicitly for the $\SLf$ duality group. We note that a
similar solution to the section condition exists for the case of $E_6$ \cite{Hohm:2013pua}.

The reduction
to IIB is achieved when the generalised metric is taken to be the fundamental
object rather than the supergravity fields themselves. These only arise upon
choosing a specific parameterisation. If one had viewed the supergravity fields
as fundamental, one could not have obtained a reduction to IIB. This
approach allows both type II theories to appear within the same
duality manifest theory.\footnote{For T-duality, the IIA and IIB theories can be treated
equally in the generalised geometry approach
\cite{Grana:2009im,Aldazabal:2010ef,Grana:2011nb,Coimbra:2011ky} and in double
field theory \cite{Siegel:1993th,Hohm:2011nu,Hohm:2011zr}. Furthermore, they can
be unified using ${\cal{N}}=2$, $D=10$ supersymmetric double field theory
\cite{Jeon:2012hp} where there is no distinction between IIA and IIB thanks to
the doubling of the local Lorentz groups,
$\mbox{Spin}(1,9)\rightarrow\mbox{Spin}(1,9)_{L}\times\mbox{Spin}(1,9)_{R }$.
This supersymmetric theory is unique while its solutions can be classified into
IIA and IIB, which are therefore unified.} 

We note that IIB appears here in a similar spirit to how it arises in F-theory. While in F-theory one introduces an extra two-dimensional space to geometrise the IIB S-duality, we are dealing with a larger U-duality group and so will use a larger extended space. As in F-theory there will be no dependence on the extra coordinates, but they are still crucial for realising the duality manifestly. Our extra coordinates will be in 
one-to-one correspondence with the wrapping modes of the branes of the IIB theory.

That it should be possible to treat M-theory and IIB in this similar way has been previously suggested by
the $E_{11}$ framework of West et al. 
\cite{West:2001as,Schnakenburg:2001he,West:2002jj,West:2003fc,Kleinschmidt:2003mf,West:2004st,
Kleinschmidt:2004rg,
Riccioni:2007au,
Riccioni:2007ni,Riccioni:2009hi,Riccioni:2009xr,West:2010ev, Rocen:2010bk} in which it is argued that both eleven-dimensional supergravity and the ten-dimensional IIA and
IIB supergravities can be understood as non-linear realisations of $E_{11}$. In
particular by
identifying the generators of the algebra corresponding to IIA and IIB fields, a
relation between the two supergravities appears directly at the ten-dimensional
level \cite{West:2004st,Riccioni:2007au}. The approach we will take in this
paper allows us to realise this idea concretely at the level
of the action.

For the sake of simplicity, we will work solely with the $\SLf$ duality
manifest theory \cite{Berman:2010is,Berman:2011cg,Berman:2011jh,Berman:2013eva,
Godazgar:2013rja,  Park:2013gaj}, originally developed for a
four-dimensional truncation of eleven-dimensional
supergravity.\footnote{Historically, $\SLf$ is viewed as the duality group that
controls the scalar sector arising from a seven-dimensional truncation of
eleven-dimensional supergravity. One assumes there is no dependence on the
four dimensions in which the $\SLf$ can be seen to act. The convention in
the duality manifest theories is different: here the truncation is taken so
that no fields depend on the seven directions and because of the introduction of
extended coordinates, the $\SLf$ symmetry becomes manifest.}
We will show how in this theory a three-dimensional
truncation of type IIB supergravity arises naturally, alongside a four-dimensional truncation of
eleven-dimensional supergravity (which can be further reduced to a truncation of type IIA
supergravity). 

We begin by briefly reviewing the key ingredients of the duality manifest field
theory of $\SLf$ in section \ref{seSL5}. In section \ref{seIIB} we then
show that the section condition has a new type of solution which cannot be 
transformed by means of an $\SLf$ duality into the M-theory solution
previously studied in the literature. Our new solution thus defines a distinct
corner of the duality manifest theory which, as we will show, corresponds to the type IIB theory.
In section \ref{seTime} we discuss what happens when one includes timelike
directions. We find not only
the usual IIB theory but also the IIB$^*$ theory
~\cite{Hull:1998vg,Hull:1998ym}. We give some concluding comments in
section \ref{seDiscuss}.

\section{The duality manifest $\mathrm{SL}(5)$ theory} \label{seSL5}

The $\SLf$ theory we use was originally formulated in
\cite{Berman:2010is} by embedding a four-dimensional truncation of
eleven-dimensional supergravity into an extended space. We will however present
a ``top-down'' approach in order to emphasise that we view the
generalised metric as fundamental and not the supergravity fields which only
arise through a specific parameterisation. This is crucial in order to obtain
the IIB theory in section \ref{seIIB}. Note that for now we take our spaces to
be Euclidean for convenience, and will deal with timelike directions in section
\ref{seTime}.

\subsection{The action, its symmetries and the section condition}
To construct an $\SLf$ duality manifest action, one uses a ten-dimensional
coordinate space such that the coordinates, $x^{ab} = -x^{ba}$, lie in the
antisymmetric representation of $\SLf$. Here $a, b = 1, \ldots, 5$ are
fundamental $\SLf$
indices. A $10\times10$ generalised metric
\cite{Duff:1990hn,Hull:2007zu,Berman:2010is} acts on this coordinate space and
parameterises the coset $\mathbb{R}^+ \!\times \SLf / \mathrm{SO}(5)$ in which
the bosonic supergravity fields we are interested in are known to live. By
restricting to $\SLf$ dualities we have implicitly performed a truncation from
some larger duality group, and this truncation gives rise to an
extra scalar
\cite{Coimbra:2011ky,Coimbra:2012af,Aldazabal:2013mya, Park:2013gaj} corresponding to the $\mathbb{R}^+$ factor.\footnote{We view this $\mathbb{R}^+$ factor as arising from a truncation of a larger duality group - alternatively, it can be thought of as a scaling symmetry as in \cite{Coimbra:2011ky}.} We discuss this in appendix
\ref{seWeighting}.

As a consequence of the coset condition the generalised metric of the $\SLf$ theory can be decomposed in terms of a
``little metric'' $m_{ab}$ \cite{Berman:2011cg}. This is a symmetric $5
\times 5$ matrix which is a rank two tensor under
$\SLf$ U-dualities. 

We can use this little metric to write the duality manifest action controlling the dynamics of this theory 
\cite{Berman:2010is,Park:2013gaj}
\begin{equation}
 \begin{split}
  S &= \int_\Sigma |m|^{-1} \left( - \frac{1}{8} m^{ab} m^{a^\prime b^\prime}
\partial_{a a^\prime} m^{c d} \partial_{b b^\prime} m_{c d} + \frac{1}{2} m^{ab}
m^{a^\prime b^\prime} \partial_{aa^\prime} m^{cd} \partial_{cb^\prime} m_{bd}
\right.\\
& \quad \left. +\frac{1}{2} \partial_{aa^\prime} m^{ab} \partial_{bb^\prime}
m^{a^\prime
b^\prime} + \frac{3}{8}  m^{ab}  m^{a^\prime b^\prime}
\partial_{aa^\prime}\ln |m| \, \partial_{bb^\prime} \ln |m| - 2 m^{a^\prime
b^\prime} \partial_{aa^\prime} m^{ab} \partial_{bb^\prime} \ln|m| \right.
\\
& \quad \left. + m^{a^\prime
b^\prime}\partial_{aa^\prime}\partial_{bb^\prime} m^{ab} - m^{ab}
m^{a^\prime b^\prime} \partial_{a a^\prime} \partial_{bb^\prime} \ln |m| \right)
\,. \label{eqSL5action}
 \end{split}
\end{equation}
Here $\Sigma$ is some lower-dimensional section of the full ten-dimensional
theory, and $m \equiv \det m_{ab}$ is used to define the $\SLf$ singlet integral measure,
$|m|^{-1}$~\cite{Park:2013gaj}.

The infinitesimal symmetries of this action are generalised diffeomorphisms, given by the generalised
Lie derivative \cite{Berman:2011cg,Berman:2012vc} 
\begin{equation}
 \mathcal{L}_{\xi} m_{ab} = \frac{1}{2} \xi^{cd} \partial_{cd} m_{ab} -
\frac{1}{2} m_{ab} \partial_{cd} \xi^{cd}+m_{cb} \partial_{ad} \xi^{cd} + m_{ac}
\partial_{bd} \xi^{cd}  \,.
\label{eqGenDiffeo}
\end{equation}
Unlike in conventional Riemannian geometry, the algebra of this generalised Lie
derivative does not automatically close. To obtain closure and a consistent
theory, one must impose a constraint known as the section condition 
\cite{Berman:2011cg}
\begin{equation}
 \partial_{[ab}
\partial_{cd]} \Phi(x) = 0 \,, \qquad  \partial_{[ab} \Phi(x) \partial_{cd]}
\Phi'(x) = 0
\,,
\label{eqSectionCondition}
\end{equation}
for any fields $\Phi(x)$, $\Phi'(x)$ in the theory. 

The action of \eqref{eqSL5action} is the most general two-derivative action
involving $m_{ab}$ that is invariant under these local symmetries up to the
section condition. It is related to the action in \cite{Berman:2010is} by an
integration by parts. It can also be given a ``U-geometric'' interpretation
\cite{Park:2013gaj} as a scalar curvature of generalised covariant derivatives,
(\ref{ScR}) (see also \cite{Aldazabal:2013mya,Cederwall:2013naa} for other
U-duality groups).

\subsection{M-theory section}

The solution of the section condition \eqref{eqSectionCondition}
which gives rise to a four-dimensional truncation of eleven-dimensional supergravity is  \cite{Berman:2010is} 
\begin{equation}
 \partial_{\alpha\beta} = 0\,, \quad \quad \partial_{\alpha5} \neq 0 \,,
\label{eqIIASection}
\end{equation}
where $\alpha, \beta = 1, \ldots, 4$. To see this, we use a convenient
parameterisation of $m_{ab}$ encoding the
metric, three-form and an extra scalar as \cite{Berman:2011cg,Park:2013gaj}
\begin{equation}
 m_{ab} = \left( \begin{array}{cc}
                  \frac{g_{\alpha\beta}}{\sqrt{|g|}} & v_\alpha \\
		  v_\beta & \sqrt{|g|} \left(e^\phi + v^\alpha v_\alpha\right)
                 \end{array} \right) \,. \label{eqMetric} 
\end{equation}
Here we have dualised the three-form field by
\be
v^\alpha = \frac{1}{3!} \epsilon^{\alpha\beta\gamma\delta}
C_{\beta\gamma\delta} \,,
\ee
where $\epsilon^{\alpha\beta\gamma\delta}$ is the four-dimensional Levi-Civita tensor,
$\epsilon^{1234} = |g|^{-1/2}$. The scalar degree of freedom, $\phi$, is needed
for the truncation from eleven dimensions to be duality covariant. Without this
scalar the action of $\SLf$ on $m_{ab}$ does not close \cite{Park:2013gaj}. One
can understand $\phi$ as coming from the truncation of the other seven
directions by constructing the generalised metric
as a non-linear realisation of $E_{11}$
\cite{Berman:2011jh} as we will discuss in appendix \ref{seWeighting}. Let us
stress that $m_{ab}$ is the fundamental object, not the supergravity fields,
$g_{\alpha\beta}, C_{\alpha\beta\gamma}, \phi$, themselves. They appear
here upon choosing a parameterisation for the M-theory section. As we show in
section \ref{seIIB}, the reduction to IIB uses a different parameterisation.

Using the solution of the section condition \eqref{eqIIASection} and the parameterisation \eqref{eqMetric} the
action reduces to \cite{Berman:2011cg,Park:2013gaj}
\begin{equation}
 S_{4d} = - \int d^4x~ e^{-2\phi} \sqrt{|g|} \left( R - \frac{1}{48} e^{-\phi}
F_{\alpha\beta\gamma\delta} F^{\alpha\beta\gamma\delta} + \frac{5}{2} \partial_\alpha \phi
\partial^\alpha \phi
\right) \,, \label{eqIIAAction}
\end{equation}
where $F_{\alpha\beta\gamma\delta}$ is the field strength of the three-form
potential,
\begin{equation}
F_{\alpha \beta \gamma \delta} = 4 \partial_{[\alpha} C_{\beta \gamma \delta]} \,.
\end{equation}
This action is a truncation of the eleven-dimensional supergravity action with the
scalar field related to the warping of the other seven dimensions. One could
introduce a further isometry and perform a Kaluza-Klein reduction of the theory
to a three-dimensional truncation of IIA supergravity.



\section{Type IIB Supergravity from the Duality Manifest Action} \label{seIIB}

In this section, we will exhibit a new solution to the section condition
allowing one to reduce to IIB supergravity. Thus, the duality manifest theory
with its extended coordinate space
allows one to treat M-theory/IIA and IIB in a unified manner. 

\subsection{IIB section}

This is a novel \emph{three-dimensional
section} which up to $\SLf$ transformations can be written as
 \begin{equation}
  \partial_{\mu i} = 0 \,, \quad \partial_{ij} = 0 \,, \label{eqIIBsect}
 \end{equation}
 where $\mu, \nu = 1, 2, 3$ and $i, j = 4, 5$. Thus the solution only
depends on $x^{12}, x^{23}, x^{31}$.

This section and the M-theory section \eqref{eqIIASection} are inequivalent, i.e. they cannot be transformed into
each other by $\SLf$. A way of seeing this is to note that 
$\partial_{[ab} \Phi \partial_{c][d} \Phi' \partial_{ef]} \Phi'' = 0$ for
the M-theory section, but $\partial_{[ab} \Phi \partial_{c][d} \Phi'
\partial_{ef]} \Phi'' \neq 0$ for the IIB section, where $\Phi,\Phi',\Phi''$ are
any three
fields in the theory. As 
$\partial_{[ab} \Phi \partial_{c][d} \Phi' \partial_{ef]} \Phi''$ transforms as
an $\SLf$ tensor, the inequivalence of the two sections follows. 

\subsection{Reduction to IIB supergravity}

We now show that the new section \eqref{eqIIBsect} should indeed be labelled as
a ``IIB section''.
Here $\mu, \nu = 1, 2, 3$ are three-dimensional spacetime indices and $i,
j = 4,
5$ are $\mathrm{SL}(2)$ S-duality indices. Our fields depend only on the coordinates
$x^{\mu \nu}$. To interpret these more naturally as spacetime coordinates we dualise
\begin{equation}
 \tilde{x}_{\mu} \equiv \frac{1}{2} \eta_{\mu\nu\rho} x^{\nu\rho} \,, \quad
\tilde{\partial}^\mu \equiv  \frac{1}{2} \eta^{\mu\nu\rho} \partial_{\nu\rho} \,,
\end{equation}
where $\eta_{123} = \eta^{123} = 1$ is the three-dimensional Levi-Civita tensor
density. 

We now parameterise the little metric $m_{ab}$
to reflect the split defined by this section:
\begin{equation}
 m_{ab} = \begin{pmatrix}
\sqrt{|\tilde{g}|} \left(\tilde{g}_{\mu\nu} +
e^{\tphi} \tilde{v}^k{}_{\mu} \tilde{v}_{k\nu} \right) & e^{\tphi}\tilde{v}_{j\mu} \\
e^{\tphi} \tilde{v}_{i\nu} & \frac{1}{\sqrt{|\tilde{g}|}}~e^{\tphi} \tilde{\cM}_{ij}
\end{pmatrix}  \,. \label{eqIIBMetric} 
\end{equation}
This parameterisation is by no means unique, but it provides a convenient way to make
a connection with type IIB supergravity.

All the fields appearing in \eqref{eqIIBMetric} can be understood as
``dual fields'' in the sense of \cite{Malek:2012pw,Malek:2013sp,
 Andriot:2011uh, Andriot:2012wx,Andriot:2012an,Andriot:2013xca,
Blumenhagen:2012pc,Blumenhagen:2012nk,Blumenhagen:2012nt,Blumenhagen:2013aia,
Blumenhagen:2013hva}. To be consistent with the
\emph{lower} index of the dual coordinates, $\tilde{x}_{\mu}$, the spacetime
metric is $\tilde g^{\mu \nu}$ with \emph{upper} indices. The determinant is
defined to be $\tilde g = \det \tg^{\mu \nu}$. The Kalb-Ramond and Ramond-Ramond
2-forms combine into an $\mathrm{SL}(2)$ doublet $\tilde C^{i \mu \nu}$, which
appears in the little metric in dualised form:
\be 
\tilde{v}^i{}_\mu = \frac{1}{2} \tilde \epsilon_{\mu\nu\rho} \tilde C^{i\nu\rho}\,,
\ee
where $\tilde \epsilon_{\mu \nu \rho}$ is the three-dimensional Levi-Civita
tensor with
$\tilde \epsilon_{123} = |\tilde g|^{-1/2}$.

In addition, we have a symmetric $2\times 2$ unit determinant matrix $\tilde{\cM}_{ij}$ which we use to raise and lower
$\mathrm{SL}(2)$ indices. This matrix can be built out of the Ramond-Ramond
scalar $\tilde C^{(0)}$ and the
string dilaton $\tilde \varphi$. A conventional parameterisation
is e.g. \cite{Polchinski:1998rr}
\be
\tilde{\cM}_{ij} = \frac{1}{\mathrm{Im}\, \tau} \begin{pmatrix} | \tau |^2 & \mathrm{Re}\,\tau \\
\mathrm{Re}\,\tau & 1 \end{pmatrix}\,, \quad \tau = \tilde C^{(0)} + i
e^{-\tilde
\varphi} \,.
\label{eqSL2M}
\ee
Finally, as in the M-theory reduction (\ref{eqMetric}),  we have an extra scalar
$\tphi$, resulting from the truncation of the
other seven dimensions, as discussed in appendix \ref{seWeighting}. Once again, this extra scalar
is necessary for the action of $\SLf$ to close \cite{Park:2013gaj}. 

The generalised Lie derivative \eqref{eqGenDiffeo} of the little metric now gives us the
expected gauge symmetries of the IIB fields. We have
\begin{equation}
 \mathcal{L}_{\xi} m_{ab} = \frac{1}{2} \xi^{cd} \partial_{cd} m_{ab} -
\frac{1}{2} m_{ab} \partial_{cd} \xi^{cd}+m_{cb} \partial_{ad} \xi^{cd} + m_{ac}
\partial_{bd} \xi^{cd} \,,
\label{eqLittlemGld}
\end{equation}
where the parameter of generalised diffeomorphisms can be decomposed into IIB language as
\be
\xi^{ab} \rightarrow ( \xi^{\mu \nu} , \xi^{i \mu}, \xi^{i j} ) \,.
\ee
For the IIB parameterisation we may dualise $\xi^{\mu \nu}$ to $\tilde \xi_\mu$ and take this to
parameterise three-dimensional diffeomorphisms. We also interpret
$\xi^{i \mu} \equiv \lambda^{i \mu}$ as corresponding to gauge transformations of the two-forms,
while the final component $\xi^{ij}$ actually drops out of the generalised Lie derivative. With this
understanding, the generalised Lie derivative \eqref{eqLittlemGld} decomposes to give
\begin{equation}
 \begin{split}
 \delta \tphi &= \txi_{\lambda} \tpartial^{\lambda} \tphi = \cL_{\txi}
\tphi\,, \\
 \delta \tilde{\cM}_{ij} &= \txi_{\lambda} \tpartial^{\lambda} \tilde{\cM}_{ij} = \cL_{\txi}
\tilde{\cM}_{ij}\,, \\
 \delta \tC^{i\mu\nu} &= \txi_{\lambda} \tpartial^{\lambda} \tC^{i\mu\nu}
+ \tpartial^{\mu} \txi_{\lambda} \tC^{i\lambda\nu} + \tpartial^{\nu}
\txi_{\lambda} \tC^{i\mu\lambda} + \tpartial^{\mu} \tlambda^{i\nu} -
\tpartial^{\nu} \tlambda^{i\mu} = \cL_{\txi} \tC^{i\mu\nu} + 2\tpartial^{[\mu}
\lambda^{|i|\nu]} \,,\\
 \delta \tg^{\mu\nu} &= \txi_{\lambda} \tpartial^{\lambda} \tg^{\mu\nu} +
 \tpartial^{\mu} \txi_{\lambda} \tg^{\lambda\nu} + \tpartial^{\nu}
 \txi_{\lambda} \tg^{\mu\lambda} = \cL_{\txi } \tg^{\mu\nu}\,,
\label{eqIIBDiffeo}
 \end{split}
\end{equation}
where
\begin{equation}
 \cL_{\txi} V^\mu \equiv \txi_\nu \tpartial^\nu V^\mu + V^\nu \tpartial^\mu
\txi_\nu \label{eqDualDiffeo}
\end{equation}
is the Lie derivative for dual diffeomorphisms
\cite{Andriot:2012an,Blumenhagen:2012nk,Blumenhagen:2012nt,Blumenhagen:2013aia}.

Having understood the field content and symmetries corresponding to this parameterisation, we are now
in a position to work out what our action is. Using the section \eqref{eqIIBsect} and the parameterisation
\eqref{eqIIBMetric} the duality manifest action \eqref{eqSL5action} reduces to\footnote{We found the symbolic
computing system Cadabra \cite{Peeters:2007wn, Cadabra} helpful in calculating this
reduction, as well as those appearing later in the paper.}
\begin{equation}
 S_{3d} = \int d^3 \tx \sqrt{|\tg|} e^{-2\tphi} \left( - \tR -
 \frac{1}{4} \tpartial^{\mu}\tilde{\cM}_{i j} \tpartial_{\mu}\tilde{\cM}^{ij}
+ \frac{1}{12} e^{\tphi} \tH^{i\mu\nu\rho} \tH_{i\mu\nu\rho}  - \frac{9}{2}
\tpartial^\mu
\tphi \,\tpartial_\mu  \tphi\right) \,,
\end{equation}
up to total derivatives. This is the expected truncated IIB action, written in terms of the dual fields and involving the extra
scalar $\tphi$ which is a relic of the truncation. The field strengths
appearing here are
\begin{equation}
\tH^{i\mu\nu\rho} = 3 \tpartial^{[\mu} \tC^{|i|\nu\rho]} \,,
\end{equation}
which are clearly invariant under the gauge symmetry of $\tC^{i\mu\nu}$
from \eqref{eqIIBDiffeo}. The unit determinant matrix $\tilde{\cM}_{ij}$ contains the dilaton $\tilde\varphi$ and RR
zero-form $\tilde C^{(0)}$, as for instance given in \eqref{eqSL2M}.
All the 
$\mathrm{SL}(2)$ indices are contracted with $\tilde{\cM}_{ij}$,
making the $\mathrm{SL}(2)$ S-duality symmetry of the IIB theory
manifest. 

Finally, note that in the dual notation, the Ricci scalar $\tR$ is defined with respect to $\tg^{\mu\nu}$, i.e. in terms
of the Levi-Civita connection $\tilde\Gamma^{\mu \nu}{}_\rho = \frac{1}{2} \tg_{\rho \lambda} (
\tpartial^\mu \tg^{\nu \lambda} + \tpartial^\nu \tg^{\mu \lambda} - \tpartial^\lambda \tg^{\mu \nu})$ and the Riemann tensor $\tilde R_\mu{}^{\nu \rho \sigma} =
2 \tpartial^{[\rho} \tilde\Gamma^{\sigma] \nu}{}_\mu + 2 \tilde\Gamma^{ [\rho|
\lambda}{}_{\mu} \tilde\Gamma^{|\sigma] \nu}{}_{\lambda}$ by $\tR = \tg_{\mu
\nu} \tilde R_{\rho}{}^{\mu \rho \nu}$. These have been
further studied in \cite{Andriot:2011uh,
Andriot:2012wx,Andriot:2012an,Andriot:2013xca,
Blumenhagen:2012nk,Blumenhagen:2012nt,Blumenhagen:2013aia}.

We can also write the action in Einstein frame by rescaling $\tilde g^{\mu
\nu}_E = e^{-4\tphi} \tg^{\mu
\nu}$ to obtain the action 
\be
S_{3d} = \int d^3 \tx \sqrt{|\tilde g_E|} \left( - \tR(\tg_E) 
- \frac{1}{4} \tpartial^{\mu}\tilde{\cM}_{i j} \tpartial_{\mu}\tilde{\cM}^{ij}
+ \frac{1}{12} e^{-7\tphi}  \tH^{i\mu \nu \rho} \tH_{i \mu
\nu \rho} + \frac{7}{2} \tpartial^\mu \tphi \,\tpartial_\mu \tphi \right) \,,
\label{eqEIIB1}
\ee 
where all spacetime contractions are now with $\tg^{\mu \nu}_E$.


\section{Lorentzian M/IIB  and M$^*$/IIB$^*$ theories}
\label{seTime}
We now consider the $\SLf$ theory relevant for Lorentzian spacetimes. The
generalised metric then parameterises the coset
space $\mathbb{R}^+ \!\times\SLf/\mathrm{SO}(3,2)$ \cite{Hull:1998br}. The
overall signature of $m_{ab}$ is $\left(-,+,+,+,-\right)$ but there
is a choice of how to distribute this. Each choice defines different versions of
the
theory. These correspond not only to the familiar eleven- and
ten-dimensional supergravities, but also to variants of these theories with unusual signatures and
kinetic terms coming with the wrong sign, which were first introduced in \cite{Hull:1998ym,
Hull:1998vg}, and have been studied in the context of $E_{11}$ and its decompositions in 
\cite{
Keurentjes:2004bv,Keurentjes:2004xx, deBuyl:2005it, Cook:2005wj
}.

Firstly, for the M-theory section, the possible parameterisations we should now
use are 
\cite{Park:2013gaj}
\begin{equation}
 m_{ab} = \left( \begin{array}{cc}
                  \frac{g_{\alpha\beta}}{\sqrt{|g|}} & v_\alpha \\
		  v_\beta & \sqrt{|g|} \left(\pm e^{\phi} + v^\alpha v_\alpha\right)
                 \end{array} \right) \,, \label{eqLMetricT}
\end{equation}
where the choice of sign in $\pm e^{\phi}$ depends on the signature of the metric $g_{\alpha\beta}$.
There are two possibilities:

\vspace{1em}
\noindent \textbf{Lorentzian M-theory:} the signature of $g_{\alpha\beta}$ is
$\left(-,+,+,+\right)$ and we have to use $-e^{\phi}$ as the scalar term in
\eqref{eqLMetricT}. The action in
Einstein frame is
 \begin{equation}
  S_{4d} = \int d^4x \sqrt{|g_E|} \left( R(g_E)- \frac{1}{48} e^{-7\phi}
F_{\alpha\beta\gamma\delta} F^{\alpha\beta\gamma\delta} - \frac{7}{2}
\partial_\alpha \phi \,\partial^\alpha \phi \right) \,,
 \end{equation}
where the Einstein frame metric is related to the one appearing in the little metric
\eqref{eqLMetricT} by $(g_E)_{\alpha\beta} = e^{-2 \phi} g_{\alpha \beta}$, and we are ignoring total
derivatives. This is of course  the known truncation of the standard eleven-dimensional supergravity \cite{Berman:2010is,Park:2013gaj}. 
 
\vspace{1em}
\noindent \textbf{M$^*$-theory:} the signature of $g_{\alpha\beta}$ is
$\left(-,-,+,+\right)$ and we have $+e^{\phi}$. The action in Einstein frame
is
 \begin{equation}
  S_{4d} = \int d^4x \sqrt{|g_E|} \left( -R(g_E) + \frac{1}{48} e^{-7\phi}
F_{\alpha\beta\gamma\delta} F^{\alpha\beta\gamma\delta} + \frac{7}{2}
\partial_\alpha \phi \,\partial^\alpha \phi \right) \,.
 \end{equation}
 The M$^*$-theory has previously been studied in terms of timelike
dualities in M-theory \cite{Hull:1998ym}.

\vspace{1em}
For the three-dimensional IIB section we may again use the parameterisation
\eqref{eqIIBMetric}, except now there are three possible ways of distributing the signature choices
between the spacetime metric $\tg^{\mu \nu}$ and the $\mathrm{SL}(2)$ scalar matrix
$\tilde{\cM}_{ij}$. These are:

\vspace{1em}
\noindent\textbf{Lorentzian IIB theory:} here the signature of $\tg^{\mu\nu}$ is
$\left(+, -,-\right)$ while $\tilde{\cM}_{ij}$ has signature $\left(+, +\right)$. We
obtain the action of the IIB theory in Einstein frame:
 \begin{equation}
  S_{3d} = \int d^3\tx \sqrt{|\tg_E|} \left( -\tR(\tg_E) - \frac{1}{4}
\tpartial_\mu \tilde{\cM}_{ij} \tpartial^\mu \tilde{\cM}^{ij} + \frac{1}{12} e^{-7\tphi}
\tH^i{}_{\mu\nu\rho} \tH^{j\mu\nu\rho} \tilde{\cM}_{ij} + \frac{7}{2} 
\tpartial_\mu \tphi\, \partial^\mu \tphi \right) \,.
 \end{equation}
 Note that $- \tR$ is the expected Ricci scalar term for the mostly negative
metric (see appendix \ref{seRicciConv} for our conventions), and so all the terms in this action
come with the expected right signs to give us a truncation of IIB supergravity
theory.

\vspace{1em}
\noindent \textbf{IIB$^*$ theory:} the signature of $\tg^{\mu\nu}$ is
$\left( -, +, + \right)$ while $\tilde{\cM}_{ij}$ has signature $\left( -, + \right)$ (and so
parameterises the coset $\mathrm{SL}(2)/\mathrm{SO}(1,1)$ rather than
$\mathrm{SL}(2)/\mathrm{SO}(2)$). The Einstein frame
action is
 \begin{equation}
  S_{3d} = \int d^3 \tx \sqrt{|\tg_E} \left( \tR(\tg_E) + \frac{1}{4}
\tpartial_\mu \tilde{\cM}_{ij} \tpartial^\mu \tilde{\cM}^{ij} + \frac{1}{12} e^{-7\tphi}
\tH^i{}_{\mu\nu\rho} \tH^{j\mu\nu\rho} \tilde{\cM}_{ij} - \frac{7}{2} 
\tpartial_\mu \tphi\, \partial^\mu \tphi \right) \,.
 \end{equation}
We note that one of the scalars of
$\tilde{\cM}_{ij}$ and one of the two-forms $\tC^{i\mu\nu}$ have kinetic terms with the
wrong signs. This corresponds to the type IIB$^*$ theory
\cite{Hull:1998ym,Hull:1998vg} which has previously been argued to arise by acting with a
timelike
duality on IIA.

\vspace{1em}
\noindent \textbf{Euclidean IIB theory:} the signature of $\tg^{\mu\nu}$
is $\left(+,
+, +\right)$ while $\tilde{\cM}_{ij}$ has signature $\left(-,-\right)$. The action in
Einstein frame becomes
 \begin{equation}
  S_{3d} = \int d^3 \tx \sqrt{\tg_E} \left( - \tR(\tg_E) - \frac{1}{4}
\tpartial_\mu \tilde{\cM}_{ij} \tpartial^\mu \tilde{\cM}^{ij} + \frac{1}{12} e^{-7\tphi}
\tH^i{}_{\mu\nu\rho} \tH^{j\mu\nu\rho} \tilde{\cM}_{ij} + \frac{7}{2}
\tpartial_\mu \tphi \,\partial^\mu \tphi \right) \,.
 \end{equation}
Note that this Euclidean IIB theory is different from \eqref{eqEIIB1} because
$\tilde{\cM}_{ij}$ has the opposite signature to the one considered there.

\vspace{1em}
\noindent This exhausts all the different possibilities. 

\vspace{1em}
The M$^*$-theory and IIB$^*$ theory were originally considered in
\cite{Hull:1998ym,
Hull:1998vg}. There
it was argued that one obtains IIB$^*$ by starting with IIA and applying a T-duality in the timelike
direction. Similarly, there is a IIA$^*$ related by a timelike T-duality to the usual
IIB theory. In both these theories one finds RR fields whose kinetic terms come with a wrong sign.
One can further view M$^*$-theory as being an eleven-dimensional theory with \emph{two} timelike
directions - compactifying on one of these timelike directions produces the IIA$^*$ theory.

In the duality covariant picture, we see these theories arising on the same footing,
corresponding to different possible choices for the parameterisation of the
generalised metric. However, they cannot be related to each other by dualities
\cite{Malek:2013sp}.


\section{Discussion} \label{seDiscuss}

In this paper we showed that the $\SLf$ duality manifest
theory contains both M-theory and IIB as different solutions of its section
condition. These solutions cannot be related by $\SLf$ transformations, and
therefore represent inequivalent sectors of the theory. In order to find both
M-theory and IIB one has to view the generalised metric as fundamental
and not the supergravity fields themselves. As a result the duality manifest
theory treats the type II theories on the same footing.

By considering the different possible choices of spacetime signature allowed by the coset structure
$\SLf/\mathrm{SO}(3,2)$ we were able to find reductions to variants of eleven-dimensional
supergravity and IIB supergravity. These variants include the normal theories as
well as the M$^*$-theory, which has two timelike directions, and the IIB$^*$
theory, for which the kinetic terms of some fields have the wrong sign. These
theories are disconnected subsectors of the full extended theory, and one cannot
transform between them by dualities.

In this paper, we considered only the duality group $\SLf$ and so obtained three-
and four-dimensional truncations of the ten- and eleven-dimensional supergravities.
The work presented here should be extended to the other duality groups
$\textrm{SO}(5,5)$, $E_6$ and $E_7$ for which all the ingredients needed for a
similar analysis are known \cite{Berman:2011jh,Berman:2012vc}. Indeed, the existence of an inequivalent
IIB section for $E_6$ has already been mentioned in \cite{Hohm:2013pua}.
Eventually, one may
wish to address this question using the full mechanics of $E_{11}$ where
independent evidence
\cite{West:2001as,Schnakenburg:2001he,West:2002jj,West:2003fc,Kleinschmidt:2003mf,West:2004st,
Riccioni:2007au,
Riccioni:2007ni,Riccioni:2009hi,Riccioni:2009xr,West:2010ev, Rocen:2010bk}
has shown that
the full ten-dimensional IIA and IIB supergravities should be contained on an
equal footing, just as in ${\cal{N}}=2$ $D=10$ supersymmetric double field
theory \cite{Jeon:2012hp} for T-duality.  

As the  section constraint  is   equivalent to gauging the extended coordinate
space~\cite{Park:2013mpa,Lee:2013hma},  the choice of a  section corresponds to
a gauge fixing. In the usual  Yang-Mills theories,  different gauge fixings are 
BRST equivalent.  It would  be interesting to further investigate the
BRST-cohomological  aspect of the two $\SLf$ inequivalent sections that we
considered here.

\section*{Acknowledgements} 
We wish to thank Malcolm Perry, Alasdair Routh and Yoonji Suh for helpful discussions.
We also thank Henning Samtleben for drawing our attention to a related work. 
JHP deeply acknowledges the hospitality
and encouragement from Hugh Osborn at DAMTP, Cambridge during his sabbatical visit. JHP is supported by the National Research Foundation of Korea (NRF) grant funded by the
Korea government (MSIP) with the Grant    No. 2012R1A2A2A02046739 and   No.
2013R1A1A1A05005747. CB is supported by the STFC, the Cambridge Home and EU
Scholarship Scheme and St John's College, Cambridge. 

\newpage 
\appendix
\section{Non-linear Realisations and the Generalised Metric}
\label{seWeighting}

By appropriately decomposing the eleven-dimensional U-duality group $E_{11}$ \cite{Berman:2011jh}, one can construct generalised metrics for  different duality groups. In the case we are concerned with in this paper, we decompose $E_{11}$ to $\SLf \times \mathrm{GL}(7)$. This can be done in two ways: one corresponding to M-theory and the other to type IIB. In both cases the generalised metric parameterises 
the coset space $(\mathrm{SL}(5) \times \mathrm{GL}(7))/ (\mathrm{SO}(5) \times
\mathrm{SO}(7)) $ (modulo assignment of the timelike direction in spacetime) and
can then be constructed as a non-linear realisation of the duality group. One
finds \cite{Berman:2011jh, Malek:2012pw}
\begin{equation} \label{eqE11}
 \mathcal{H}_{MN}  = |\tg_7|^{-1/2} |\tg|^{-1/2}
\begin{pmatrix}
\tM_{ab,cd} & 0 \\						
0 & (\tg_7)_{AB}
\end{pmatrix} \,.		
\end{equation}
Here the indices $M$, $N$ run over the full $7+10$ dimensional extended space, while $A,B$ are indices in the seven dimensions which will be truncated in what follows and $a,b, \dots$ as before are fundamental $\SLf$ indices. 
The $10\times 10$ matrix $\tM_{ab,cd}$ carries pairs of antisymmetric such indices and is what is actually normally referred to as
the generalised metric. It will have some particular parameterisation in terms
of spacetime fields and in particular its determinant is $\det \tM = |\tg|^{-2}$, with $|\tg|$ being the determinant of the metric of the three- or four-dimensional
truncation of spacetime which we are going to restrict to. Finally $(\tg_7)_{AB}$
is the metric for the other seven directions.\footnote{ 
In the rest of the paper, we used $\tg$ to refer to the IIB spacetime metric, to
stress that it was given in
dual form. However in this appendix $\tg$ refers to the spacetime metric
appearing in the $E_{11}$ generalised metric \eqref{eqE11}.}
 We are assuming a block diagonal decomposition of the full spacetime metric,
and also that all components of form fields are zero when any index corresponds
to one of these seven directions.

The full generalised metric $\mathcal{H}_{MN}$ is a
weighted $\mathrm{SL}(5)$ tensor, and can be used to directly construct a
Lagrangian density. However, we want to truncate to the ten-dimensional extended
space. 
This truncation involves using only the upper left block of the generalised metric $\mathcal{H}_{MN}$. We shall from now on refer to this block as the generalised metric, and denote it
\be
\bar M_{ab,cd} = |\tg_7|^{-1/2} |\tg|^{-1/2} \tM_{ab,cd} \,.
\ee
Neither this generalised metric nor the generalised metric $\tM_{ab,cd}$ is an
element of $\SLf$, as can be seen by considering their determinants. One can
define an $\SLf$ generalised metric $\hat{M}_{ab,cd}$ with $\det \hat{M}=1$ by
$\hat{M}_{ab,cd} = |\tg|^{1/5} \tM_{ab,cd}$. However, it is not possible to
construct the dynamics of the theory using $\hat{M}_{ab,cd}$, as discussed in
\cite{Berman:2011jh}.

We are therefore forced to use generalised metrics which are not truly
elements
of $\SLf$. However these generalised metrics will transform correctly under
duality transformations by virtue of the factor of the determinant of the metric
in the truncated directions appearing outside. It is this $|\tg_7|^{-1/2}$
factor which corresponds to including an extra scalar in \cite{Park:2013gaj} and
also in our treatment. 

The precise way in which one includes this scalar is a matter of taste. One could take a double field theory-esque approach, and define a new scalar density $e^{-2d}$ by
\be
\bar M_{ab,cd}= |\tg_7|^{-1/2} |\tg|^{-1/2}  \tM_{ab,cd} = e^{2d}
\hat{M}_{ab,cd} \,,
\ee
where one finds that $e^{2d} = |\tg_7|^{-1/2} |\tg|^{-7/10}$. Alternatively, the
$|\tg_7|$ factors can be absorbed into field redefinitions. 

Explicitly, suppose one decomposes the generalised metric $\tM_{ab,cd}$ in terms
of a little metric $\tm_{ab}$, so that similarly $\bar M_{ab,cd}$ can be written
in terms of a little metric $\bar m_{ab}$ given by
\be
\bar m_{ab} = |\tg_7|^{-1/4} |\tg|^{-1/4} \tm_{ab} \,.
\ee
We know that the determinant of $\tm_{ab}$ is $|\tg|^{-1/2}$. We define a new
generalised metric in which the factor of $|\tg_7|^{-1/4}$ has been absorbed by
\be
\bar m_{ab} = |\det m |^{1/2} m_{ab} 
\ee
One sees that 
$m_{ab}$ is unweighted with respect to the generalised Lie derivative and is given by 
$m_{ab} = |\tg_7|^{-1/14} \tm_{ab}$. 

For both the M-theory and IIB parameterisations it is then a simple matter to
determine the precise rescalings of the spacetime fields that make up $\tm$: for
instance, for the M-theory case this involves equating 
\be
m_{ab} =
 \begin{pmatrix}
 \frac{g_{\alpha\beta}}{\sqrt{|g|}} & v_\alpha \\
 v_\beta & \sqrt{|g|} \left(\pm e^\phi + v^\alpha v_\alpha\right)
\end{pmatrix} 
=
|\tg_7|^{-1/14} \begin{pmatrix}
 \frac{\tg_{\alpha\beta}}{\sqrt{|\tg|}} & \tv_\alpha \\
 \tv_\beta & \sqrt{|\tg|} \left(\pm 1 + \tv^\alpha \tv_\alpha\right) 
\end{pmatrix} \,.
\ee
Here $g$ is the spacetime metric that is used for the M-theory truncation
throughout the paper. Furthermore, we find here that $e^\phi =
|\tg_7|^{-3/14}$ while for the IIB generalised metric \eqref{eqIIBMetric} we
have $e^{\tphi} = |\tg_7|^{-2/7}$.
The little metric in the form corresponding to $m_{ab}$ with the scalar $e^\phi$
present was first used in \cite{Park:2013gaj}, and it is this which we use
throughout the present work.

\section{Conventions and useful formulae\label{seRicciConv}}

\subsection{Conventions for spacetime curvature}

Our spacetime Riemann curvature tensor is defined in terms of the Christoffel symbols $\Gamma_{\mu
\nu}{}^\rho$ of the Levi-Civita connection
\be
R^{\mu}{}_{\nu \rho \sigma} = \partial_\rho \Gamma_{\sigma \nu}{}^\mu -
\partial_\sigma \Gamma_{\rho \nu}{}^\mu 
+ \Gamma_{\rho \lambda}{}^{\mu} \Gamma_{\sigma \nu}{}^{\lambda} 
- \Gamma_{\sigma \lambda}{}^{\mu} \Gamma_{\rho \nu}{}^{\lambda} \,,
\label{eq:Riemann}
\ee
and the Ricci scalar is 
\be
R = g^{\mu \nu} R^{\rho}{}_{\mu \rho \nu} \,.
\ee
With this definition the Einstein-Hilbert Lagrangian for a Lorentzian metric of mostly plus signature is then $+R$, while that for a
Lorentzian metric of mostly minus signature is $-R$. 

For the dual fields used in the type IIB reduction, these definitions are the same except with the
positions of all indices flipped, i.e. $\tR$ is the Ricci scalar defined with respect to
$\tg^{\mu\nu}$, using the Levi-Civita connection $\tilde\Gamma^{\mu \nu}{}_\rho = \frac{1}{2}
\tg_{\rho \lambda} ( \tpartial^\mu \tg^{\nu \lambda} + \tpartial^\nu \tg^{\mu \lambda} - \tpartial^\lambda \tg^{\mu \nu})$ and the Riemann tensor $\tilde R_\mu{}^{\nu \rho \sigma} =
2 \tpartial^{[\rho} \tilde\Gamma^{\sigma] \nu}{}_\mu + 2 \tilde\Gamma^{ [\rho| \lambda}{}_{\mu}
\tilde\Gamma^{|\sigma] \nu}{}_{\lambda}$. This has been
further studied in \cite{Andriot:2011uh,
Andriot:2012wx,Andriot:2012an,Andriot:2013xca,
Blumenhagen:2012nk,Blumenhagen:2012nt,Blumenhagen:2013aia}.

If $g_{\mu \nu}$ and $\bar g_{\mu \nu} = \Omega g_{\mu \nu}$ are two $D$-dimensional metrics
related by a conformal scaling then
\be
\bar R = \Omega^{-1} \left( R - (D-1) \nabla^2 \ln \Omega - \frac{1}{4} (D-2)(D-1) ( \nabla \ln
\Omega)^2\right)\,.
\ee

\subsection{Alternative form of the duality manifest action}

Following  \cite{Park:2013gaj}, the $\SLf$ U-duality covariant action~(\ref{eqSL5action}) can be
equivalently formulated as
\be
\displaystyle{S=\int_{\Sigma} |m|^{-1}{\cal R}\,,}
\label{ScR}
\ee
where ${\cal R}$ is the  invariant  scalar curvature given by the generalized metric and its derivatives,
\be
{{\cal R}}=\partial_{ab}(2A^{acb}{}_{c}-A^{abc}{}_{c})+\half
A_{abcd}A^{abcd}-2A_{abcd}A^{acbd}-\textstyle{\frac{1}{4}}
A_{abc}{}^{c}A^{abd}{}_{d}-2A_{cab}{}^{c}A^{abd}{}_{d}+2A_{cab}{}^{c}A^{dba}{}_{
d}\,,
\ee
while $A_{abcd}$ is the  simplest choice for a  metric-compatible,
semi-covariant derivative connection,
\be
A_{abcd}=A_{[ab](cd)}=\half m_{cd}m^{ef}\partial_{ab}m_{ef}-\half\partial_{ab}m_{cd}\,.
\ee
\bibliographystyle{JHEP}
\bibliography{BibliographyL}

\end{document}